%
%
\magnification=\magstep1
%
%
\baselineskip = 1.5\baselineskip 
\parskip=2pt plus 1pt
\abovedisplayskip=18pt plus 4.5pt minus 13.5pt
\abovedisplayshortskip=0pt plus 4.5pt
\belowdisplayskip=\abovedisplayskip
\belowdisplayshortskip=10.5pt plus 4.5pt minus 4.5pt
\smallskipamount= 4.5pt plus 1.5pt minus 1.5pt
\medskipamount=9pt plus 3pt minus 3pt
\bigskipamount=18pt plus 6pt minus 6pt
%
%
\newif\ifnonlin \nonlinfalse
\hsize=210truemm \vsize=297truemm
\ifnonlin
  \advance\hsize by -6truecm
  \advance\hoffset by -1truein \advance\hoffset by 4truecm
\else
  \advance\hsize by -2truein
\fi
\advance\vsize by -2truein \advance\vsize by -\baselineskip
\overfullrule=0pt
%
%

\font\tenfrak=eufm10
\font\tif = cmcsc10 scaled\magstep2
\font\sc = cmcsc10
\font\secfont = cmbx12 \def\Bf{\secfont\relax}
\def\frak#1{\tenfrak #1\relax}
\def\Cal#1{{\cal #1}}
%
%
\let\o\omega \let\t\theta

\def\H{\Cal H} \def\I{{\cal I}}
\def\D{\Cal D}
\def\A{\Cal A} \def\An{\A_n}
\def\C{\Cal C} \def\Cn{\C_n}
\def\I{\Cal I} \def\M{\Cal M}
\def\R{{\bf R}}
\def\P{{\bf P}} 
\def\RP{{\bf RP}}
%
%
\def\SL{\hbox{\rm SL}} \def\SLn{\SL(n,\R)}
\def\sln{\hbox{\frak{sl}}(n,\R)}
\def\sL#1{\hbox{\frak{sl}}(#1,\R)}
\def\GL{\hbox{\rm GL}}
\def\gl(n, \R){C^{\infty}(S^1, gl (n, \R))}
%
%
\def\res{\mathop{{\rm res}}}
\def\pr{\mathop{{\rm pr}}\nolimits}
\def\Id{\mathop{{\rm Id}}\nolimits}
%
%
\def\frac#1#2{{#1\over#2}}
\def\dd#1{\frac\partial{\partial#1}}
\def\ddi{\dd{\phi_i}} \def\ddj{\dd{\phi_j}}
\def\pij{\phi_i^{(j)}}
\def\ddij{\dd{\pij}}
%
%
\def\bo#1{\hbox{\bf #1}} \def\v{\bo v} \def\w{\bo w}
%
%
\def\bin#1#2{{#1\choose #2}}
%
%
 
%

\def\forall{\qquad\hbox{\rm for all}\quad}
%
%
\def\adots{\mathinner{\mkern2mu\raise1pt\hbox{.}\mkern2mu
\raise4pt\hbox{.}\mkern2mu\raise7pt\hbox{.}\mkern1mu}}
%
%
\def\eqtag{\eqno(\the\sectionno.\the\formulano)}
\def\Eq#1{%
\global\advance\formulano by 1
\eqtag
\expandafter\ifx\csname zzz#1\endcsname\relax
 \global\expandafter
 \edef\csname zzz#1\endcsname{\the\sectionno.\the\formulano}%
\else\immediate\write0{Warning: equation #1 already defined!}%
 \edef\csname zzz#1\endcsname{{\bf #1}}\fi
}
\def\eqn#1{\expandafter\ifx\csname zzz#1\endcsname\relax
 \immediate\write0{Equation #1 not provided!}{\bf #1}%
\else\csname zzz#1\endcsname\fi} 
\def\eq#1{{\rm (\eqn{#1})}}
%
%
\def\neqalign#1{\def\eqtag{&(\the\sectionno.\the\formulano)}
\eqalignno{#1}}
%
%
\newcount\sectionno
\newcount\subsectionno
\newcount\formulano
\newcount\theoremno
\sectionno=0 \subsectionno=0
\def\Section#1 #2. {%
\theoremno=0\formulano=0 \subsectionno=0
\advance\sectionno by1
\goodbreak\removelastskip\bigskip\noindent
{\Bf \the\sectionno.~#2}\par\nobreak\medskip\noindent
\expandafter\ifx\csname szzz#1\endcsname\relax
 \expandafter\edef\csname szzz#1\endcsname{\the\sectionno}%
\else\immediate\writeout{Warning: Section #1 already defined!}%
 \expandafter\edef\csname szzz#1\endcsname{{\bf #1}}\fi
}
\def\section#1{Section \sec #1}
\def\sec#1{\expandafter\ifx\csname szzz#1\endcsname\relax
\immediate\write0{Section #1 not provided!}{\bf #1}%
\else\csname szzz#1\endcsname\fi}
\def\subsection#1. {\advance\subsectionno by1
\removelastskip\medskip\noindent{\it
\the\sectionno.\the\subsectionno.~#1}\par\nobreak
\smallskip\noindent}
%
%
\def\th#1{\expandafter\ifx\csname tzzz#1\endcsname\relax
\immediate\write0{Theorem #1 not provided!}{\bf #1}%
\else\csname tzzz#1\endcsname\fi}
\def\proclaim#1#2.{%
\expandafter\ifx\csname tzzz#1\endcsname\relax
 \global\advance\theoremno by1
 \smallskip\noindent{\bf \the\sectionno.\the\theoremno~#2.}%
 \global\expandafter
 \edef\csname tzzz#1\endcsname{\the\sectionno.\the\theoremno}\enspace
 \begingroup\sl
\else\immediate\write0{Warnig: Theorem #1 already defined!}\fi
}
\def\endproclaim{\endgroup\smallskip\noindent}
\def\proof{{\sc Proof.}\quad}
\def\qed{\quad{\it Q.E.D.}}
%
%
\def\TheDate{\ifcase\month\or January\or February\or March\or 
April\or May\or June\or July\or
August\or September\or October\or November\or December\fi 
\space\number\day, \number\year}
%
%
\newbox\refbox
\newdimen\refdimen
\setbox\refbox=\hbox{[OST]}
\refdimen=\wd\refbox \advance\refdimen by .5em
\def\rf#1{[#1]}
\def\[#1]{\item{\hbox to \refdimen{[#1]\hfill}}}
%
%
%
%
\rightline{\sc UCM Preprint--3/96}
\bigskip
\centerline{\tif Invariant Differential Equations }
\smallskip
\centerline{\tif and }
\smallskip
\centerline{\tif the Adler--Gel'fand--Dikii Bracket}
\bigskip
\centerline
{\it Artemio Gonz\'alez-L\'opez\kern 1.5pt${}^1$}
\centerline{\it Rafael Hern\'andez Heredero\kern 1pt${}^1$}
\centerline{\it Gloria Mar\'\i{} Beffa\kern 1pt${}^2$}
\medskip
{\parskip = 3truept \baselineskip = .83333\baselineskip
\item{${}^1$}Depto.~de F\'\i sica Te\'orica II,
Facultad de Ciencias F\'\i sicas, Universidad Complutense, 28040 Madrid, Spain.
Supported in part by DGICYT under grant no.~PB92--0197.\par
\item{${}^2$}Dept.~of Mathematics, University of Wisconsin, Madison,
WI 53706, USA.\par\vskip-\parskip\hang
Current address: Depto.~de Matem\'atica Aplicada, Facultad
de Matem\'aticas, Universidad Complutense, 28040 Madrid, Spain.\par}
\medskip
\centerline{\TheDate}
\bigskip
\noindent {\bf Abstract}\quad
In this paper we find an explicit formula for the most general
vector evolution of curves on $\RP^{n-1}$ invariant under the
projective action of $\SLn$. When this formula is applied to the
projectivization of solution curves of scalar Lax operators with periodic
coefficients, one obtains a corresponding evolution in the space of such
operators. We conjecture that this evolution is identical to the second KdV
Hamiltonian evolution under appropriate conditions. 
These conditions give a Hamiltonian interpretation of general
vector differential invariants for the projective action of $\SLn$, namely,
the $\SLn$ invariant evolution can be written so that a general vector
differential invariant corresponds to the Hamiltonian pseudo-differential
operator. We find common coordinates and
simplify  both evolutions so that one can attempt to prove the equivalence for
arbitrary $n$.
\medskip
\noindent {\it Short title:}\quad Invariant Equations and the AGD Bracket.
\vfill\break
\Section intro Introduction.
In this paper we try to answer the following question: let $L(t,\theta)$ be a
family of scalar differential operators with periodic coefficients following an
evolution (in $t$) which is Hamiltonian with respect to the second KdV
Hamiltonian structure or Adler--Gel'fand--Dikii bracket. Consider a
family of solution curves $\xi(t,\theta)$ associated to $L(t,\theta)$. Is there
a simple way to describe the evolution of
$\xi(t,\theta)$? The importance of studying the space of solutions of $L$ was
pointed out by Wilson in \rf{Wi}. These curves are also used to  provide a
discrete invariant of the Poisson bracket, one of the two invariants which
classify the symplectic leaves, \rf{OK}. The answer to this question could,
of course, offer an alternative definition of the bracket.

Here, we aim to show that the evolution of the solution curve is important
by itself. In fact, we will see that the evolution of its projectivization
$\phi(t,\theta)$ corresponds to the most general evolution of curves on real
$(n-1)$-dimensional projective space $\RP^{n-1}$ which is
invariant under the projective action of $\SLn$. The study of the explicit form
of this kind of equations lies naturally within the scope of the classic subject
of Klein geometries and differential and geometric invariants, which had its
high point in the last century before the appearance of Cartan's approach to
differential geometry. Recently Olver et al., \rf{OST}, used this theory to
characterize all scalar evolution equations invariant under the action of a
subgroup of the projective group in the plane, a problem of interest in
the theory of image processing.
%
%
Following Olver's approach,
we will write explicitly the most general (vector) evolution of curves on
$\RP^{n-1}$ of the form 
$$
\phi_t = F(\phi, \phi_{\theta}, \phi_{\theta\theta}, \dots )
$$
which is invariant under the $\SLn$ projective action.
%
%
We conjecture that this invariant equation corresponds to the
Adler--Gel'fand--Dikii evolution under the duality described  below. This
correspondence can be shown to be true for many fixed values of $n$, but we
haven't succeeded in proving the general case, which is considerably more
involved. We will guide the reader in simplifying the proof in the general
case, so that he or she can attempt to prove the conjecture for any particular
value of $n$.

Denote by $\An$ the infinite-dimensional manifold of
scalar differential operators (or Lax operators) with $T$-periodic 
coefficients of
the form $$
L = \frac{d^{n}}{d\t^n} +  u_{n-2}
\frac{d^{n-2}}{d\t^{n-2}} + \cdots +  u_1 \frac{d}{d\t} + u_0 ,
\Eq{ln}
$$
and let $\xi_L = (\xi_1, \dots, \xi_n)$ be a solution curve associated to 
$L$ the Wronskian of whose components equals one. Due to the 
periodicity of the
coefficients of $L$, there exists a matrix $M_L \in \SL(n, \R)$, called 
the {\it
monodromy} of $L$, such that
$$\xi_L(\t + T) = M_L 
\xi_L(\t),\qquad\hbox{\rm for all }\t \in \R.$$
($M_L$ is defined by the Floquet matrix of the 
differential
equation.) This same property holds for its (non-degenerate) 
projection on the
$n-1$ sphere   $S^{n-1}$ ($\hat\xi_L = \frac{\xi_L}{|\xi_L|}$,
where $|\,\cdot\,|$ represents the norm on $\R^n$), and is also 
shared by the projective coordinates of this projection, whenever we consider 
the actions of $\SL(n,\R)$ on the sphere and on projective space, respectively.
Observe
that the  monodromy is not completely determined by the operator $L$, but by 
its
solution 
curves.
Namely, if one chooses a different solution curve, its monodromy 
won't be
equal to $M_L$ in general, but it will be the conjugate of $M_L$ by 
an element
of $\GL(n, \R)$. That is, $L$ only determines the conjugation class of the 
monodromy.
Of course, this problem does not exist once the solution curve has been fixed.

Conversely, let $\phi: \R \to \R\P^{n-1}$ be a curve on $\R\P^{n-1}$.
Assume that the curve $\phi$ is {\it non-degenerate and right-hand oriented},
%
%
that is the Wronskian determinant of the components of its derivative $\phi'$
is positive. (This is equivalent to the Wronskian of the components of $(\phi,
1)$ being positive; for example, the curve would be convex and right-hand
oriented in  the case $n=3$.) Assume also  that
$\phi$ satisfies the following {\it monodromy property:}
$$
\phi(\t + T) = (M
\phi)(\t),\qquad\hbox{\rm for all }\t \in \R,
\Eq{monphi}
$$
for a given $M \in \SL(n, 
\R)$. Here
$M\phi$ represents the usual action of $\SL(n, \R)$ on $\R\P^{n-1}$, 
induced by the action of $\SL(n, 
\R)$ on $\R^n$.
One can associate to $\phi$ a differential operator of the form \eq{ln} 
in the
following manner: We lift $\phi$ to a curve on $\R^n$, say to
$f(\t)(1,\phi)$. We choose the factor $f$ so that the 
Wronskian of
the components of the new curve equals 1. There is a unique choice of $f$ 
with such a property (up to perhaps a sign), namely
$$
f = W(1,\phi_1, \dots, \phi_{n-1})^{-1/n}
=W(\phi_1', \dots,\phi_{n-1}')^{-1/n},
$$ 
where $\phi 
= (\phi_1,
\dots, \phi_{n-1})$ and $W$
represents the Wronskian determinant.

It is not very hard to see that the coordinate functions
of the lifted curve are solutions of a unique differential operator of 
the
form \eq{ln}. Such an operator defines an equation for an unknown $y$ 
of the form 
$$
\left|\matrix{y&f\phi_0&\dots&f\phi_{n-1}\cr
y' & (f \phi_0)'& \dots & (f 
\phi_{n-1})'\cr
\vdots& \vdots & \ddots & \vdots \cr 
y^{(n)} & (f \phi_0)^{(n)} & \dots & (f
\phi_{n-1})^{(n)}}\right| = 0;\qquad\phi_0=1,\quad' = \frac{d}{d\t}.
\Eq{deq}
$$
Equation \eq{deq} can be written in the usual manner as a system of
first order differential equations $\frac{d X }{d\t} = N X$, where 
$$N =
\pmatrix{0 & 1 & 0 & \dots  & 0 \cr 0 & 0 & 1 & \dots & 0 \cr 
\vdots & \vdots &
\ddots & \ddots & \vdots \cr 0 & 0 & \dots & 0 & 1 \cr -u_0 & -u_1 
& \dots &
-u_{n-2} & 0}$$
and $X$ is a fundamental matrix solution 
associated to the
differential equation \eq{deq}. From this formulation and the 
monodromy condition
it is trivial to see that $N = \frac{d X}{d\t} X^{-1}$ is a periodic 
matrix
and so are the coefficients of the operator defining \eq{deq}.

A short comment is due at this point: if $M$ is the monodromy matrix associated
to $\phi$, for even $n$ the monodromy matrix associated to 
$L$ could be either $M$ or $-M$, depending on whether the first
component of $M(1,\phi)$ is positive or negative. Hence, it would be more 
correct to talk about the action of ${\rm P}\SLn$, the space obtained from
$\SLn$ by  identifying
$M$ and $-M$. Since this choice makes no difference in what follows,
we will keep $\SLn$ for the sake of simplicity.

From these descriptions we get two parallel pictures, 
one in the 
manifold $\An$ and another one in the space $\Cn$ of non-degenerate curves 
on $\RP^{n-1}$. In the next sections, we will study how the
Adler--Gel'fand--Dikii Hamiltonian evolution in $\An$ is related to the
$\SLn$ invariant evolution in $\Cn$. We will explicitly show that, for low
values of
$n$, both evolutions are equivalent under the above identification, and we
will conjecture that this is actually true for all values of $n$.

\Section agd The evolution equations on $\An$.
\subsection The Adler--Gel'fand--Dikii bracket.
We start by describing one of the Hamiltonian evolutions on the manifold 
$\An$, the well known Adler--Gel'fand--Dikii bracket, or second KdV 
Hamiltonian Structure.

Given a linear functional $\cal H$ on $\An$, one can associate to it a 
pseudo-differential operator
$$H = \sum_{i=1}^n h_i \partial^{-i},\qquad\partial 
= \frac{d}{d\t},
$$
such that 
$$\H(L) = \int_{S^1} \res (HL)\,d\t,$$
where $\res $ selects 
the coefficient of $\partial^{-1}$ and is called the {\it residue} of the 
pseudo-differential operator (see \rf{A} or \rf{GD}). To any $\H$ we 
can associate a (Hamiltonian) vector field $V_H$ defined as 
$$V_H (L) = (L H)_+ L- L(HL)_+,$$
where by $({}\cdot{})_+$ we denote the non-negative (or 
differential) part of the operator. The map $H \to V_H$ is 
a {\it structure map} defining a Poisson bracket on the manifold 
$\An$. If $\hat\ell$ is the matrix of differential operators
defining the structure map, the Poisson bracket is defined as 
$$
\{\H, \Cal F\}(L) = 
\int_{S^1}\res (\hat\ell(H) F)\,d\t, \Eq{poisson}
$$
cf.~\rf{A}, \rf{GD} or \rf{O1}. The original 
definition of the bracket was given by Adler, \rf{A}, in an attempt to 
make generalized KdV equations bi-Hamiltonian systems. Gel'fand 
and Dikii proved Jacobi's identity in \rf{GD}.
In the case $n=2$, this bracket coincides with the Lie-Poisson 
structure on the dual of the Virasoro algebra. Two other equivalent 
definitions of the original bracket were found in \rf{KW} and in \rf{DS}. 
The original definition is rather complicated, so we will explain and 
use the one in \rf{KW}.  

\subsection The Kupershmidt--Wilson bracket.
In a very interesting paper, \rf{KW}, Kupershmidt and Wilson gave an 
equivalent but rather simpler definition of the bracket \eq{poisson}.
Consider $L$ to be an operator of the form \eq{ln}. Assume that the 
operator $L$ factors into a product of first-order factors
$$
L = (\partial+y_{n-1})(\partial + y_{n-2}) \cdots (\partial + y_1)(\partial 
+ y_0),
$$
where 
$$y_k = \o^k v_1 + \o^{2k} v_2 + \cdots + \o^{(n-1)k} v_{n-1}, 
\qquad 0\le k\le n-2;\quad \o = e^{\frac{2\pi i}{n}},
\Eq{ydef}$$
and $y_{n-1} = -\sum_{i=0}^{n-2} y_i$.
The variables $v_i$, $1\le i\le n-1$, are what Kupershmidt and Wilson called
``modified" variables. Even though the factorization is not unique (and so
some reduction had to be involved in  the proof of \rf{KW}), one can find a
unique factorization once a solution curve has been fixed, as
we will see later. 

Assume that the coefficients $u_i$, $0\le i\le n-2$, of $L$
evolve  following a Hamiltonian evolution with respect to the second KdV
Hamiltonian  structure. The result in \rf{KW} then states that the
corresponding ``modified" coordinates $v_i$ evolve following a
Hamiltonian evolution with  respect
to a Poisson bracket defined by the structure map
$$
\neqalign{
\ell &= -\frac 1n \partial J ,   \Eq{vev1}\cr
\noalign{\rm where}
J &= \pmatrix{0&\dots&0&1 \cr 0 & \dots & 1&0\cr
\vdots&\adots&\adots&
\vdots\cr
1&0&\dots&0}.}
$$
That is, 
$$
\frac{Du}{Dv}\,\ell \left(\frac{Du}{Dv}\right)^{\ast} = \hat\ell,
\Eq{lhat}
$$
where 
$\displaystyle\frac{Du}{Dv} = \left(\frac{Du_i}{Dv_j}\right)$,
$$
\frac{Du_i}{Dv_j}=
\sum_{k=0}^{n-1} \frac{\partial u_i}{\partial v_j^{(k)}}\, \partial^k$$ 
being the Fr\'echet
derivative of $u_i$ with respect to $v_j$. Also, by ${}^{\ast}$ we denote the
adjoint matrix operator, the transposed of the matrix whose entries are the
adjoint operators of the entries of the original matrix. Thus, the original
Adler--Gel'fand--Dikii bracket arises from a very simple bracket defined on the
space of ``modified"  variables $v$.

Many facts are known about this Hamiltonian structure. Since it is 
Poisson (degenerate), the manifold $\An$ foliates into symplectic 
leaves, maximal submanifolds where the Hamiltonian flow always 
lies. These leaves are classified locally by the conjugation class  of 
the monodromy matrix associated to the operators lying on the leaf. 
In other words, if two operators are close and have conjugate monodromies, 
there is a Hamiltonian path joining them. There exists another 
discrete invariant that classifies the leaves globally, cf.~\rf{OK}, based on
topological properties of the projection of the solution
curves on the sphere $S^n$.

\Section iee Invariant evolution equations on $\Cn$.
%
The duality between $\An$ and $\Cn$ described in
the Introduction makes it natural to study evolution equations on the space
$\Cn$ whose associated flow leaves the Adler--Gel'fand--Dikii symplectic leaves
invariant. In other words, we are interested in partial differential equations
of the form
$$
\phi_t = F(\t,\phi, \phi_{\t}, \phi_{\t\t},
\dots),\qquad\phi:\R^2\to\RP^{n-1},
\Eq{ev}
$$
for a function
$\phi(\t,t)$, with the property that, if the initial condition has a 
monodromy property \eq{monphi}, then every solution
$\phi(\cdot,t)$ of \eq{ev} has also a monodromy property, and the conjugation
class of the monodromy matrix is independent of $t$. The simplest evolution
equations having this
property are those of the form \eq{ev} with $F$ independent of $\t$
which are also invariant under the standard projective action of $\SLn$ on the
dependent variables $\phi=(\phi_1,\dots,\phi_{n-1})$. In other words, we are
dealing with equations of the form
$$
\phi_t = F(\phi, \phi_{\t}, \phi_{\t\t}, \dots),
\qquad\phi:\R^2\to\RP^{n-1},
\Eq{evi}
$$
such that whenever $\phi(\t,t)$ is a solution of \eq{evi} so is
$(M\phi)(\t,t)$, for all $M\in\SLn$. To see that the monodromy class of the
solutions $\phi(\cdot,t)$ of an equation \eq{evi}
invariant  under the action of
$\SLn$ is indeed preserved under the evolution, note that \eq{evi}
is also invariant under translations of the independent variable
$\t$. Hence, if the initial condition $\phi(\cdot,0)$ of \eq{evi} has a matrix
$M\in \SL(n, \R)$ as monodromy, and we consider a different curve in the flow 
$\phi(\cdot,t)$, we have that $\phi(\t-T,t)$ is also a solution.
If \eq{evi} is $\SLn$-invariant, $M\phi(\t-T,t)$ will also be a
solution  of \eq{evi}. Applying uniqueness of solutions of \eq{evi} (whenever
possible),
$M\phi(\t-T,t) =
\phi(\t,t)$, so that $\phi(\cdot,t)$ has the same monodromy as
$\phi(\cdot,0)$. If there is no uniqueness of solutions, both Hamiltonian and 
invariant evolutions are obviously much more complicated; we won't deal
with those cases in this paper.

\noindent{\it Remark:} note that the evolution associated to an
$\SLn$ invariant equation \eq{evi} preserves exactly the monodromy (not just
the monodromy class) of its solutions.
\medskip
In this paper we conjecture that the Adler--Gel'fand--Dikii evolution on
$\An$ and the $\SLn$ invariant evolution \eq{evi} on
$\Cn$ are  identical
under the identification described in the Introduction, provided that the
coefficients of  the
Hamiltonian $H$ (the  pseudo-differential operator describing the differential 
of
the  functional $\H$) are equal to a  vector
differential invariant  of
the projective action. We will find the most general $\SLn$ invariant
evolution of the form
\eq{evi}, showing then how the conjecture can be proved for a number of
values of $n$ and where the main problem lies in the proof of the general case.

The most general evolution equation of the form \eq{evi} invariant
under the projective action
$$
\phi(\t,t)\mapsto (M\phi)(\t,t)
$$
of $\SLn$ can be found using the general infinitesimal
techniques described in \rf{O1}, \rf{O2}.
%
%
First of all, the
infinitesimal generators of the projective $\SLn$ action are easily found to
be the following vector fields on $\R\times\R\times\RP^{n-1}$:
$$
\v_i = \ddi,\qquad \v_{ij} = \phi_i\ddj,\qquad
\w_i = \phi_i\sum_{j=1}^n \phi_j\ddj; \qquad 1\le i,j\le n-1.
\Eq{sln}
$$
The vector fields \eq{sln} are a basis of a realization of the Lie algebra
$\sln$. Note that all these vector fields are independent of the variables
$(\t,t)$, and they are also ``vertical", i.e, their $\t$ and $t$ components
vanish. 

If $\v=\sum_{i=1}^{n-1}\eta_i(\t,t,\phi)\ddi$ is a vertical vector field, its
{\it prolongation} is the vector field $\pr \v$ defined by
$$
\pr \v = \v + \sum_{j\ge1\atop k\ge0}\sum_{i=1}^{n-1}
\left(D_t^k\,D^j\eta_i\right)\dd{(\partial_t^k\pij)},
\Eq{prt}
$$
where $\pij=\partial^j\phi_i$, $D$ is the total derivative operator with
respect to $\t$
$$
D = \partial + \sum_{j\ge0}\sum_{i=1}^{n-1}\phi_i^{(j+1)}\ddij,
$$
and
$$ D_t = \partial_t + \sum_{j\ge0}\sum_{i=1}^{n-1}
\left(\partial_t\pij\right)\ddij
\Eq{dt}
$$
is the total derivative operator with respect to $t$. In general, the vector
field $\pr \v$ is defined on the
infinite-dimensional jet space
$J^\infty(\R\times\R,\RP^{n-1})$ with local coordinates
$\t,t,\partial_t^k\pij$ ($1\le i\le n-1;\ k,j\ge0$). However, when $\pr \v$ is
applied to a function (like $F$) independent of the coordinates
$\partial_t^k\pij$ ($k\ge1$) involving explicitly $t$-derivatives, \eq{prt}
reduces to the vector field
$$
\pr \v = \v + \sum_{j\ge1}\sum_{i=1}^{n-1}
\left(D^j\eta_i\right)\ddij,
\Eq{prol}
$$
defined on the infinite-dimensional jet space
$J^\infty\equiv J^\infty(\R,\RP^{n-1})$ with local coordinates
$\t,\pij$ ($1\le i\le n-1,j\ge0$). Following \rf{O1}, we can express 
the necessary and sufficient
condition for
\eq{evi} to be invariant under the action of $\SLn$ ``infinitesimally" as
follows:
$$
\pr \v(F) = 
D_t\eta
\left.\vrule width 0pt height 0pt depth4pt
\right|_{\phi_t=F},\forall
\v=\sum_{i=1}^{n-1}\eta_i(\phi)\ddi\in\sln.
\Eq{sym}
$$
Note that, although both
$\pr \v$, $D$ and $D_t$
are formally defined on infinite-dimensional jet spaces, in practice they
will always act on functions depending on a finite number of the local
coordinates. Finally, using the fact that $\eta$ is a function of $\phi$ only
and \eq{dt}, equation \eq{sym} can be further simplified as follows:
$$
\pr \v(F) = \frac{\partial\eta}{\partial\phi}\, F,
\Eq{sym2}
$$
where $\frac{\partial\eta}{\partial\phi}$ is the $(n-1)\times(n-1)$ matrix
with
$(i,j)$ entry
$\frac{\partial\eta_i}{\partial\phi_j}$. In other words, $F$ is a {\it
relative vector differential invariant} of the Lie algebra $\sln$ given
by \eq{sln}, whose associated {\it weight} is the matrix
$\frac{\partial\eta}{\partial\phi}$. Using standard techniques
(cf.~\rf{O2}), we can give the following characterization of the general
solution of
\eq{sym2}:
\proclaim{gensym}Theorem.
The most general solution $F$ of equation \eq{sym2} is of the form
$$
F = \mu\, \I,
$$
where the $(n-1)\times(n-1)$ matrix
$\mu=(\mu^1\; \mu^2\; \dots\; \mu^{n-1})$ is any matrix with non-vanishing
determinant and whose columns $\mu^i$ are particular solutions of
\eq{sym2}, and $\I=(\I_k)_{k=1}^{n-1}$ is an arbitrary {\rm absolute
(vector) differential invariant} of the algebra \eq{sln}, i.e.~a solution of
$$
\pr \v(\I_i) = 0, \forall \v\in\sln,\quad i=1,\dots,n-1.
$$
\endproclaim

The problem of calculating the most general absolute differential invariant
$\I$ of a given Lie algebra of vector fields is a classical one,
\rf{H}, \rf{L},
\rf{T}, whose solution in a modern formulation can be found in
%
%
\rf{O2}. The
general result asserts that their exist $n$ functionally independent
differential {\it fundamental invariants} $J_0$, $J_1$,\dots,$J_{n-1}$, such
that any differential invariant is a  function of the $J_i$'s and their
``covariant derivatives" $\D^k J_i$, where
$\D=(DJ_0)^{-1}\,D$. Since in our case the generators \eq{sln} are
independent of $\t$, we can take $J_0=\t$, so that the operator $\D$ reduces
to $D$ in this case. Therefore, we can state the following Theorem:

\proclaim{gensln}Theorem.
The most general ($\t$-independent) absolute differential invariant of the
$\sln$ Lie algebra
\eq{sln} is a function of $n-1$ fundamental differential invariants
$J_i(\phi,\dots,\phi^{(m)})$ and their total derivatives with respect to $\t$.
\endproclaim

For $n=2$, it is straightforward to compute the fundamental
$\sL2$ invariant $J_1$. The result is the classical {\it Schwartzian
derivative} $S(\phi)$ of $\phi$:
$$
J_1 = \frac{\phi'''}{\phi'}-\frac32 \frac{\phi''^2}{\phi'^2}. \Eq{Sch}
$$
In this case, the matrix $\frac{\partial\eta}{\partial\phi}$ is just a
function, which makes a simple matter to find a particular vector
differential invariant of weight $\frac{\partial\eta}{\partial\phi}$. The
simplest such invariant is $\phi'\equiv\phi_\t$; therefore, Theorem
\th{gensln} implies the following:
\proclaim{genslnev}Theorem.
For $n=2$, the most general evolution equation \eq{evi} invariant under the
projective action of $\SLn$ is
$$
\phi_t = \phi_\t\,\I(S,DS,\dots,D^l S),
$$
where $S$ is the Schwartzian derivative of $\phi(\cdot,t)$, and $\I$ is an
arbitrary (smooth) function.
\endproclaim

Even for the case $n=3$, it is not an easy matter to find the $n-1$
fundamental differential invariants of \eq{sln} and a particular
matrix relative differential invariant of weight
$\frac{\partial\eta}{\partial\phi}$ from scratch. Fortunately, however, the
differential invariants of the projective action of $\SLn$ have been the
object of considerable study in classical projective differential geometry,
\rf{W}. From this viewpoint, the differential invariants of a projective
curve describe the properties of the curve invariant under the group of
motions of projective space, or in other words the properties of
the curve independent of the particular system of projective coordinates
used to represent it. An intrinsic description of a projective curve must
therefore be done in terms of its $\sln$ differential invariants. It is not
hard to see (as we will explain in the following section) that the 
coefficients of the 
operator $L$ defined by a projective curve $\phi$ as in \eq{deq} are a set of
functionally independent differential invariants. Obviously, they determine
the curve up to a projective transformation; this was already known to
Wilczynski,
\rf{W}, and it is a generalization of the well known result in Euclidean
geometry that the curvatures of a curve in Euclidean space, expressed as
functions of the Euclidean-invariant arclength, uniquely characterize the
curve up to an Euclidean motion. We shall explain in the following sections
how this equivalence between fundamental differential invariants and
coefficients of the operator
$L$ is the key to the duality of evolutions.

\Section ef The explicit formula for the $\SLn$ invariant
evolution. 
In this section we will describe a complete set of independent differential
invariants for the projective action of $\SLn$, and we will give the explicit
expression of the relative invariant
\eq{sym2} with the required weight, for arbitrary $n$. The
complete set of differential invariants was already found by \rf{W} and is
precisely given by the coefficients of the operator $L$ determined by the
curve $\phi$, as mentioned in the Introduction.

\proclaim{inv}Theorem.  
Let $\phi$ be a non-degenerate and right-hand oriented curve on $\RP^{n-1}$,
and let 
$$
L = \frac{d^{n}}{d\t^n} +  u_{n-2}
\frac{d^{n-2}}{d\t^{n-2}} + \cdots +  u_1 \frac{d}{d\t} + u_0 
$$
be the differential operator determined by $\phi$ through the relation
{\rm \eq{deq}}. Then the coefficients $u_i$,  $0\le i\le n-2$, form a complete
set of  functionally independent differential invariants for the $\SLn$ action
on
$\RP^{n-1}$. \endproclaim
\proof Using the form of equation \eq{deq} one can easily see that the
coefficient of $\frac{d^k}{d\t^k}$ is given by
$
u_k = - \Delta_k  
$, 
where $\Delta_k$ is the determinant
obtained from the Wronskian determinant $W(f,f\phi_1, \dots, f\phi_{n-1}) =
1$ when we substitute the $(k+1)^{\rm th}$ row by the $n^{\rm th}$ derivative
row
$\bigl(f^{(n)},(f\phi_1)^{(n)}, \dots, (f\phi_{n-1})^{(n)}\bigr)$. Thus, the
coefficients of $L$ are functions of the components of the curve $\phi$ and
their derivatives.
From this it follows that the coefficients of the operator $L$ are
functionally independent functions. Indeed, if there were a functional
relation among these functions one could choose an operator whose
coefficients did not satisfy this relation. The projectivization $\phi$ of
the solution curve of such an operator would then have coefficients
$u_k(\phi)$, $k= 0, \dots, n-2$, not satisfying the functional relation,
and we would get a contradiction.

The coefficients $u_i$ are easily shown to be invariants. Indeed, let $M \in
\SLn$ and let $M \phi$ be the image of the curve $\phi$ under the
projective action of $M$. If we lift $\phi$ to a solution curve of $L$, say
$(f,f\phi)$, and we also lift the curve $M \phi$, we see that the latter is
simply
$M\cdot(f,f\phi)$ (the dot denoting matrix multiplication). Since
$M\cdot(f,f\phi)$ represents a non-degenerate linear combination of the
solution curve
$(f,f\phi)$, both lifted curves  are solutions of the same operator and
hence
$u_k(\phi) = u_k(M\phi)$ for all
$k$.\qed

\smallskip
Next, we will find the explicit expression for $n$ independent relative
vector invariants, solutions of \eq{sym2}
for all vector fields $\v=\sum_{i=1}^{n-1}\eta_i(\phi)\,\ddi\in\sln$.
That is, we want to find a matrix
$$
\mu = (\mu^1\;   \mu^2\;   \dots\;  \mu^{n-1}) \Eq{mat}
$$
each of whose columns $\mu^i$ is a
solution of equation \eq{sym2}, and such that the determinant of $\mu$ does
not vanish.

Before going into the details of how one finds this matrix, we need
several preliminary definitions and results:

\proclaim{def}Definition. 
For $i_1,\dots,i_k\ge0$ and $1\le k\le n-1$, let us denote 
$$\eqalignno{
w_{i_1i_2\dots i_k}&=
   \left|\matrix{ \phi_1^{(i_1)} & \phi_2^{(i_1)} & \dots & \phi_k^{(i_1)} \cr
                  \phi_1^{(i_2)} & \phi_2^{(i_2)} & \dots & \phi_k^{(i_2)} \cr
                  \vdots & \vdots & \ddots & \vdots \cr
        \phi_{1}^{(i_k)} & \phi_{2}^{(i_k)} & \dots &\phi_{k}^{(i_k)}}
      \right|\cr
\noalign{\rm and}
W_{k}&=w_{12\dots k}.}
$$
We define the {\rm homogeneous variables} $q_{i_1i_2\dots i_k}$ by
$$
q_{i_1i_2\dots i_k}=\frac{w_{i_1i_2\dots i_k}}{W_k}.
$$
Finally, for $k=1,2,\dots,n$ the variables $q_n^k$  are defined as follows:
$$
q_n^k = q_{12\dots\widehat k\dots n},
$$
where the notation $\widehat k$ means that the index $k$ is to be omitted.
\endproclaim
The following statements follow easily from elementary properties of
determinants:
\proclaim{form}Lemma.\par\nobreak
\item{\rm i)} For any $k,i_1, \dots ,i_r\ge0$ and $1\le s<r\le n-1$ we have the
following identities:
$$
\openup2pt\eqalign{
q_k q_{i_1 i_2 \dots i_r}
&= q_{i_1} q_{k i_2 \dots i_r} + q_{i_2} q_{i_1 k\, i_3 \dots i_r}+ \dots
+q_{i_r} q_{i_1
\dots i_{r-1} k},\cr
q_{i_1i_2\dots i_s k} q_{i_1 i_2 \dots i_r} &= 
q_{i_1\dots i_si_{s+1}} q_{i_1 \dots i_s k\, i_{s+2}\dots i_r}\cr
&\quad{}+ 
q_{i_1\dots i_si_{s+2}} q_{i_1 \dots i_{s+1} k\, i_{s+3}\dots i_r}+
\cdots +
q_{i_1\dots i_si_r} q_{i_1 \dots i_{r-1} k }.}\Eq{form1}
$$
\item{\rm ii)}  If we define $q_m^0 = 0$ for all $m\ge2$, then the
following identity holds:
$$
q_n^k = q_{n-1}^k q_n^{n-1} - q_{n-1}^k q_{n-1}^{n-2} - (q_{n-1}^k)' +
q_{n-1}^{k-1},
\qquad 1 \le k < n.
$$
\endproclaim
Note that $q_n^n=1$ by definition. The {\it affine algebra} is the subalgebra
of the
$\sln$ algebra
\eq{sln} generated by the vector fields $\v_r$ and $\v_{rs}$, $1\le
r,s\le n-1$. The corresponding group of transformations is the affine
group, i.e., the semidirect product of the translation group with the general
linear group in the variables $(\phi_1,\dots,\phi_{n-1})$.
\proclaim{coor}Lemma. If a $\t$-independent function $\psi: J^\infty\equiv
J^\infty(\R,\RP^{n-1}) \to \R$ is invariant under the action of the affine
algebra, then $\psi$ necessarily depends only on the affine coordinates
$q_n^r$,
$r = 1, \dots, n-1$, and their derivatives.
\endproclaim
\proof
Consider the prolonged action of the affine algebra on the $k^{\rm th}$ jet
space
$J^k\equiv J^k(\R,\RP^{n-1})$, whose infinitesimal generators are the
$k^{\rm th}$ prolongations (i.e., the truncations of the prolongations
\eq{prol} at differential order $k$)
$$
\pr^{(k)} \v_r=\v_r,\qquad
\pr^{(k)} \v_{rs}=\sum_{j=0}^k \phi_r^{(j)}\,\dd{\phi_s^{(j)}}.
\qquad 1\le r,s\le n-1.
\Eq{afgen}
$$
For $k\le n-1$, at a generic point of $J^k$ the $n(n-1)$ vector
fields~\eq{afgen} span the $(k+1)(n-1)$-dimensional subspace of the tangent
space of $J^k$ whose elements are the ``vertical" vector fields (whose
component along $\dd\t$ vanishes).  By Frobenius theorem, this implies that
there are no affine differential invariants of differential order between $1$
and $n-1$, and the only zero-th order invariant is clearly (a function of)
the coordinate
$\t$. It is also immediate to check that for $k\ge n-1$ the vector fields
\eq{afgen} are linearly independent at a generic point. Hence the maximal
dimension of the span of these vector fields stabilizes for $k=n-1$. Olver's
general results, cf.~\rf{O2}, imply that the affine algebra has $n-1$
fundamental invariants of order $n$, and that an arbitrary
differential invariant can be expressed as a function of $\t$,
the fundamental invariants, and their derivatives with respect to the zero-th
order invariant $\t$. Since the
$n-1$ functions
$q_n^r$, $1\le r\le n-1$, have all differential order $n$, and are clearly
functionally independent and invariant under general affine transformations of
the variables
$(\phi_1,\dots,\phi_{n-1})$ by their definition, they can be taken as the
$n-1$ fundamental invariants.\qed
\proclaim{gen}Lemma. The variables $q_r^s$ ($r>s\ge1$) can be written
in terms of the functionally independent functions $q_k^{k-1}$ ($k\ge2$) and
their derivatives. We will call the latter functions {\rm basic homogeneous
variables.}
\endproclaim
\proof
We will prove the lemma by induction on $r-s$. For $r-s = 1$,
the lemma holds trivially. Assume now that the functions
$q_{r'}^{s'}$ with
$r'-s' < m$ can be expressed in terms of the functions
$q_k^{k-1}$ and their derivatives. Let $q_r^s$ be such that $r-s = m$. From
ii) of Lemma \th{form} we have that
$$ q_r^s = q_{r-1}^s q_r^{r-1} - q_{r-1}^s q_{r-1}^{r-2} - (q_{r-1}^s)' +
q_{r-1}^{s-1},
$$ so that by the induction hypothesis $q_r^s$ can be written in terms of the
functions
$q_k^{k-1}$ and their derivatives
if, and only
if, the same is true for $q_{r-1}^{s-1}$. Repeating this argument $s-2$ times,
we see that $q_r^s$ will be a function of the $q_k^{k-1}$'s and
their derivatives, if only if this is the case for $q_{m+1}^1$, with
$m=r-s>0$. Again from ii) in Lemma
\th{form}, we have that
$$
q_{m+1}^1 = q_m^1 q_{m+1}^m - q_m^1 q_m^{m-1} - (q_m^1)'
$$
which, by the induction hypothesis, proves the lemma.\qed

\smallskip
We are now going to make an ansatz for the matrix $\mu$. Namely, we will
look among matrices $\mu$ of the form 
$$
\mu = \Phi\,( \Id + A),
\Eq{mat2} 
$$
where 
$$
\Phi = \pmatrix{ \phi_1' & \phi_1'' & \dots & \phi_1^{(n-1)} \cr
\phi_2' & \phi_2'' & \dots & \phi_2^{(n-1)} \cr \vdots & \vdots & \ddots &
\vdots
\cr \phi_{n-1}' & \phi_{n-1}'' & \dots & \phi_{n-1}^{(n-1)}},
\Eq{Phi}
$$
$\Id$ is the identity matrix, and $A$ is a strictly upper triangular matrix
to be determined. Obviously, a matrix $\mu$ of this form will have a
non-vanishing determinant.

\proclaim{rinv}Theorem.
An invertible matrix $\mu$ of relative invariants with weight $
\frac{\partial\eta}{\partial\phi}$ is given by a matrix of the form
{\rm \eq{mat2}--\eq{Phi},} with $A = (a_i^j)$ defined by
$$
a_i^j = \cases{\hfill
 \frac{\displaystyle(-1)^{j-i}\bin ji
\vrule width 0pt height 0pt depth 11pt}{\displaystyle\bin n{j-i}
\vrule width0pt height 16pt depth 15pt}\, q_n^{n-j+i},\hfill&
\quad$i < j$\cr
\hfill0,\hfill&\quad$i \ge j$.}
\Eq{as}
$$
\endproclaim
\proof We only need to show that each one of the columns of $\mu$ is a
particular solution of equation \eq{sym2}. Assume that 
$\mu = (\mu^1\;  \dots \; 
\mu^{n-1})$ is of the form \eq{mat2}--\eq{Phi}, so that $\mu^i =
(\mu^i_j)_{j=1}^{n-1}$ is a column given by $\mu^i_j = \phi_j^{(i)} +
\sum_{k=1}^{i-1} a_k^i
\phi_j^{(k)}$. Assume also that $\v=\sum_{i=1}^{n-1}\eta_i(\phi)\ddi\in\sln$. 
We can then write equation \eq{sym2} as
$$
\pr \v (\mu^i_j) = \sum_{k = 1}^{n-1}\frac{\partial \eta_j}{\partial
\phi_k}\,\mu^i_k.
\Eq{sym3}
$$
Obviously, it suffices that \eq{sym3} hold for all the basic vector fields
\eq{sln}. We will therefore consider the following three cases: 

\smallskip
(a)\quad If $\v = \v_r = \dd{\phi_r}$, then $\pr \v = \v$ and \eq{sym3}
trivially holds, since both sides of the equality vanish.

\smallskip
(b)\quad If $\v = \v_{r s} = \phi_r \dd{\phi_s}$, then its prolongation is
given by $\pr \v_{r s}= \sum_{k \ge0} \phi_r^{(k)} 
\frac{\partial}{\partial\phi_s^{(k)}}$. Substituting in \eq{sym3}, we obtain
the equivalent equation
$$
\sum_{k=1}^{i-1} \phi_j^{(k)} \pr \v_{r s}(a_k^i) = 0,
\qquad 1\le i, j, r, s\le n-1.
$$
In matrix notation the latter equation becomes
$$
\Phi \,\pr \v_{r s}(A) = 0,\qquad r,s = 1, 2, \dots, n-1,
$$
and since $\Phi$ is invertible for all projective curves under consideration
this is equivalent to
$$
\pr \v_{r s}(A) = 0, \qquad r,s = 1, 2, \dots, n-1.  \Eq{inv1}
$$
(By $\pr \v_{r s}(A)$ we mean the matrix
obtained when we apply the vector field $\pr \v_{r s}$ to each of the
entries of the matrix $A$.)
Since the matrix $A$ in \eq{as} depends only on the affine
invariant coordinates $q_n^r$, $1\le r\le n-1$, by Lemma \th{coor} we deduce
that \eq{inv1} holds for this matrix.

\smallskip
(c)\quad If $\v = \w_r = \phi_r \sum_{k=1}^{n-1} \phi_k\, \dd{\phi_k}$, its
prolongation is given by the formula 
$$ \pr \w_r =  \sum_{j\ge0} \sum_{k = 1}^{n-1} 
(\phi_r \phi_k)^{(j)}\, \dd{\phi_k^{(j)}}.
$$
Substituting this formula into \eq{sym3}, we easily arrive at the matrix
equation 
$$
\Phi\,\pr \w_r(A) = \hat\Phi_r\,(\Id + A),
\qquad r=1, 2, \dots n-1,
$$
where
$$
(\hat\Phi_r)_j^i = 
\phi_r \phi_j^{(i)} +  \phi_j \phi_r^{(i)} - (\phi_j\phi_r)^{(i)}.
$$ 
The product $\Phi^{-1} \hat\Phi_r$ can be easily rewritten in a nice
way. In fact, the $(j, i)$ entry of this product is given by
$$
-\sum_{k=1}^{n-1} \phi_j^k \sum_{l=1}^{i-1} \bin il
\phi_k^{(l)} \phi_r^{(i-l)},
$$
where $\phi_j^k$ is the $(j,k)$ element of $\Phi^{-1}$. Now, since 
$\sum_{k=1}^{n-1} \phi_j^k\,\phi_k^{(l)} = \delta_j^l$, the $(j, i)$ entry of
the product 
$\Phi^{-1}
\hat\Phi_r$ equals zero if $j \ge i$ and $-\bin ij
\phi_r^{(i-j)}$ whenever $j < i$. Therefore, the infinitesimal
invariance condition in  case (c) is given by
$$
\pr \w_r(A) = -\Gamma_r\,(\Id + A), \qquad r= 1, 2, \dots, n-1,
\Eq{act1}
$$
where
$$
\Gamma_r = \pmatrix{0 & \bin21 \phi_r' &
\bin31 \phi_r'' & \dots & \bin{n-1}1
\phi_r^{(n-2)} \cr 0 & 0 & \bin32 \phi_r'& \dots &
\bin{n-1}2 \phi_r^{(n-3)} \cr \vdots & \vdots & \ddots
& \ddots & \vdots \cr 0 & \ddots & 0 & 0 & \bin{n-1}{n-2} \phi_r'
\cr 0 & \dots & 0 & 0 & 0 }.
$$
To complete the proof, we  only need to check that \eq{act1} is satisfied when
$A$ is given by \eq{as}.
What follows are straightforward calculations.

First of all, one can easily see that
$$
\eqalign{
\pr \w_r(w_{1 2 \dots \widehat k \dots n}) &= \sum_{{j=1 \atop j\ne k}}^n
\phi_r w_{1 2 \dots \widehat k \dots n} + \sum_{j=1}^{k-1}(-1)^{j-1}
\phi_r^{(j)} w_{0 1 \dots \widehat j \dots \widehat k \dots n}\cr
&\enspace\kern-2.15129pt
{}+ \sum_{j=k+1}^n (-1)^j \phi_r^{(j)} w_{0 1 \dots \widehat k \dots
\widehat j \dots n} + \sum_{j=k+1}^n \bin jk (-1)^{j-k+1}
\phi_r^{(j-k)} w_{1 2 
\dots \widehat j \dots n}.}
$$
Using formula \eq{form1} we obtain
$$
\phi_r q_n^k = \sum_{j=1}^{k-1} (-1)^{j-1} \phi_r^{(j)} q_{0 1 \dots
\widehat j
\dots \widehat k \dots n} + \sum_{j=k+1}^n (-1)^j \phi_r^{(j)}  q_{0 1 \dots
\widehat k \dots \widehat j \dots n},
$$
so that
$$
\pr \w_r(w_{1 2 \dots \widehat k \dots n}) = n \phi_r w_{1 2 \dots \widehat
k
\dots n}  + \sum_{j=k+1}^n \bin jk (-1)^{j-k+1} \phi_r^{(j-k)}
w_{1 2 
\dots \widehat j \dots n}.
$$
Applying Leibniz's rule we
finally obtain
$$
\pr \w_r(q_n^k) = \sum_{j=k+1}^n \bin jk(-1)^{j-k+1}
\phi_r^{(j-k)} q_n^j.
$$
If we substitute in \eq{act1} the value of $A$ given in
the Theorem and use the expression of $\pr \w_r(q_n^k)$ derived above
\eq{act1} becomes
$$
\eqalign{
&(-1)^{j-i} \frac{\bin ji}{\bin n{j-i}}
\sum_{l = n-j+i+1}^n \bin l{n-j+i} (-1)^{l-n+j-i+1} 
\phi_r^{(l-n+j-i)} q_n^l
\cr
&\quad{}= - \sum_{l=i+1}^{j} 
\bin li \phi_r^{(l-i)} (-1)^{j-l} \frac{\bin jl}{\bin n{j-l}} q_n^{n-j+l},
\qquad 1\le i< j\le n-1.
}
$$
This equation will hold provided that
$$
\frac{\bin ji\,\bin {n+l-j}{l-i}}{\bin n{j-i}}=
\frac{\bin li\,\bin jl}{\bin n{j-l}},
$$
which is indeed an identity, since both sides equal
$$
\frac{j!\, (n+l-j)!}{i!\, n!\, (l-i)!}.
$$
This concludes the proof of the Theorem.\qed

As an immediate consequence we obtain the
following corollary:

\proclaim{inveq}Corollary. The most general equation for the evolution of 
curves on $\RP^{n-1}$ which is invariant under the projective action of $\SLn$
is given by 
$$ 
\phi_t = \Phi\,(\Id + A)\, \I,
\Eq{inveq}
$$
where $\Phi$ and $A$ are given by \eq{Phi} and \eq{as}, and $\I$ is
any vector differential invariant for the action.
\endproclaim

\Section eqe The equivalence of evolutions.
\subsection The $\SL(2, \R)$ case.
We will describe the case $n = 2$ first to illustrate the procedure to
be
followed in general. In this case $\An\equiv\A_2$ is the manifold of
Hill's operators of the form
$$
\frac{d^2}{d\t^2} + u,
\Eq{lii}
$$
and $\Cn\equiv\C_2$ is the space of curves on the projective line such that
$\frac{d\phi}{d\t} = \phi_{\t} \ne 0$. By Theorem \th{genslnev}, the most
general evolution on
$\C_2$ invariant under the $\SL(2, \R)$ action is given by the equation
$$
\phi_t = \phi_{\t} \I.\Eq{invde}
$$
Here $\I$ is a differential invariant of the action, that is, a function of
$S(\phi)$ and its derivatives with respect to $\t$, where $S(\phi)$ is the
Schwartzian derivative of $\phi$ given by \eq{Sch}.
	
Given a curve $\phi$ on $\C_2$ with a monodromy $M$, there is a unique 
operator of
the form \eq{lii} such that $\xi = (\xi_1, \xi_2) =
(\phi'^{-\frac12},\phi'^{-\frac12} 
\phi)$ is its solution curve. Once the solution curve is fixed
one can
factor $L = (\partial-v)(\partial+v)$ in a unique fashion so that $(\partial 
+
v)\,\xi_1=0$ and $(\partial-v)(\partial+v)\,\xi_2=0$. More precisely, 
$$
v = -\frac{(\phi'^{-\frac12})'}{\phi'^{-\frac12}} =
\frac12\frac{\phi''}{\phi'}.
$$ 
Assume now that
$\phi$ is evolving according to equation \eq{invde}. Then, due to its
dependence on
$\phi$,  $v$ will be evolving following the equation
$$
\eqalign{
v_t &= \frac{Dv}{D\phi}(\phi' \I) = 
-\frac{D\left((\phi'^{-\frac12})'/
\phi'^{-\frac12}\right)}{D\phi}\,(\phi' \I)\cr &=
-\partial\left(\frac1{\phi'^{-\frac12}}
\frac{D(\phi'^{-\frac12})}{D\phi}\right)(\phi'\I)
=
\frac12 \partial \left(\frac1{\phi'}\partial
\right)(\phi' \I) =
\frac12\partial(\partial + 2v) \I.}
$$
On the other hand, the evolution of $v$ according to the Kupershmidt--Wilson
definition is given
by
$$
v_t = -\frac12\,\partial\,\frac{\delta \H}{\delta v} \Eq{vev2},
$$
for some Hamiltonian functional $\H$ depending on $v$ and its derivatives.

Two comments are due at this point. First of all, let $\frac{\delta
\H}{\delta v}$ be the kernel of the Fr\'echet derivative of
$\H$ with respect  to $v$,
and let $\widehat{\frac{\delta \H}{\delta u}}$ denote the corresponding kernel
with respect to the variables
$u$, expressed in terms of $v$. Then the following equality holds:
$$\frac{\delta \H}{\delta v}=
\left(\frac{Du}{Dv}\right)^{\ast}  \widehat{\frac{\delta \H}{\delta u}}.$$
The
proof of this  statement can be found in \rf{KW}, p.~420.

The second comment is as follows: notice that $\frac{\delta \H}{\delta u}$ is
a differential  invariant, since it depends on the coefficients $u$ and their
derivatives, which are themselves independent differential invariants. This was
pointed out throughout Sections 3 and 4. On the other hand, the latter result
doesn't hold for the Fr\'echet derivative with respect to $v$, since the
coefficients of the first-order factors are not invariant
with respect to the action of $\SL(n, \R)$. Thus, in order to find the
equivalence
of evolutions, we must write the Adler--Gel'fand--Dikii evolution of $v$ in
terms of the Hamiltonian
as a function of $u$. That is, the proper correspondence is
between
the $\phi$-evolution and the $u$-evolution, since the coefficients $u$ are
invariants of the $\SLn$ action. We are using the variables $v$ to simplify 
calculations,
since the original definition of the Adler--Gel'fand--Dikii bracket in terms of
the
$u$ coordinates is too complicated. These two comments are obviously valid in
the general case and not only for
$n=2$.

Returning to \eq{vev2},
we can rewrite this equation as
$$
v_t = -\frac12\partial \left(\frac{Du}{Dv}\right)^{\ast} 
\widehat{\frac{\delta \H}{\delta u}}
$$
But $ u =  -v^2+v'$ in this case, so
that
$\left(\frac{Du}{ Dv}\right)^{\ast} = -(\partial + 2v)$, and we have thus
shown that the evolution
due to the dependence of $v$ on $\phi$ is identical to the 
Adler--Gel'fand--Dikii
evolution provided that
$$\I = \frac{\delta \H}{\delta u}.$$ 

\subsection The general case.
The proof for other values of $n$ follows the same ideas that we showed in the
case $n=2$. The main practical problem is, of course, the complication of the
calculations involved. Our goal is to show that whenever a non-degenerate
right-hand oriented projective curve $\phi$ follows the evolution $\phi_t =
\Phi\,(\Id + A)\,\I$, then the corresponding coefficients of its associated
operator follow the Adler--Gel'fand--Dikii evolution provided the
vector differential invariant
$\I$ is related to $\frac{\delta\H}{\delta u}$ in a suitable way. In this
section we will simplify the problem and establish a closer connection between
both evolutions before showing where the main problem lies. In any case, using
this simplified version it is relatively easy to establish the equivalence of
both evolutions for a fixed value of
$n$.

\proclaim{modv}Proposition. A choice of modified variables $v$ can be
expressed in terms of the basic homogeneous variables as 
$ v =  \Omega^{-1} y $, where
$$ 
y_0 = \frac1n
q_n^{n-1},\quad
y_i = q_i^{i-1} - q_{i+1}^i + \frac1n q_n^{n-1}, \qquad 1\le i \le n-2,
\Eq{yq}
$$
$q^0_1 = 0$ by definition, and $\Omega$ is the
Vandermonde matrix defined by
$$
\Omega = \pmatrix{1 & 1 & \dots & 1 \cr \o & \o^2 & \dots & \o^{n-1} \cr
\vdots &\vdots &\ddots &\vdots\cr \o^{n-2} & \o^{2(n-2)}& \dots &
\o^{(n-1)(n-2)}}.
$$
\endproclaim 
\proof
It suffices to show that the we can factor $L =
\partial^n + u_{n-2} \partial^{n-2} + \cdots + u_1 \partial + u_0 =
(\partial + y_{n-1}) \cdots (\partial + y_0)$ uniquely so that the
coefficients $y$ are given by \eq{yq}. Let us lift $\phi$ to a solution of
$L$. The solution is given uniquely by $\xi = (\xi_1, \xi_2, \dots, \xi_n) =
W_{n-1}^{-1/n}(1,\phi_1, \phi_2, \dots,
\phi_{n-1})$. We choose $y$
so that $\xi_i$ is a solution of 
$$
(\partial + y_{i-1})\cdots(\partial +
y_1)(\partial + y_0)\, \xi_i = 0,\qquad i = 1, \dots, n.
$$
It is not hard to
show that there is a unique choice for $y$, namely
$$ y_i = \frac{\o'_{i-1}}{\o_{i-1}} - \frac{\o'_i}{\o_i},
\qquad i=1,2,\dots,n,
\Eq{yo}
$$ 
where 
$\o_i = W(\xi_1,\dots,\xi_{i+1})$.
Indeed, notice that $y_0+y_1+\cdots+y_{i-1}$ is the coefficient
of $\partial^{i-1}$ in $(\partial + y_{i-1})\cdots(\partial + y_1)(\partial +
y_0)$. On the other  hand, if $\xi_1, \dots, \xi_i$ are the independent
solutions of this operator, then the coefficient of $\partial^{i-1}$ is given
by
$-\o_{i-1}' /\o_{i-1}$, cf.~\eq{deq}, from which \eq{yo} easily follows. From
the form of
$\xi$ we get that 
$\o_i = W_i/
W_{n-1}^{(i+1)/n}$. Substituting in \eq{yo} we get \eq{yq} for $i\ge 1$
straightforwardly. The formula for $y_0$ is an immediate consequence of the
equation $(\partial+y_0)\,\xi_1 = (\partial+y_0)\,W_{n-1}^{-1/n}=0$, while the
relationship
$v =
\Omega^{-1} y$ is simply the definition of \eq{ydef} of $v$.\qed

We want to see next under what conditions the evolution of $v$ 
$$
v_t = \frac{Dv}{D\phi}\, \phi_t = \frac{Dv}{D\phi}\, \Phi\, (\Id + A)\, \I
\Eq{maine1}
$$
induced by the $\SLn$ invariant evolution of $\phi$ coincides with the
Adler--Gel'fand--Dikii Hamiltonian evolution
$$
v_t =
-\frac1n\,
\partial J \left(\frac{Du}{Dv}\right)^{\ast} \widehat{\frac{\delta \H}{\delta
u}},
\Eq{maine100}
$$
where
$\widehat{\frac{\delta
\H}{\delta u}}$ is defined as in the case $n=2$. We are  going to simplify
both equations before proceeding with further calculations. Using the previous
proposition, we can write \eq{maine1} as
$$
v_t = \Omega^{-1} S\, \frac{Dq}{D\phi}\, \Phi\, (\Id + A)\, \I,
$$
where
$$
S = \pmatrix{0&0&0&\dots &0&{1/ n}\cr
{-1}&0&0&\dots &0&{1/ n}\cr
1&{-1}&0&\dots &0&{1/ n}\cr
0&\ddots &\ddots &\ddots &\vdots &\vdots \cr
\vdots &\ddots &1&{-1}&0&{1/ n}\cr
0&\dots &0&1&{-1}&{1/ n}\cr
}
$$
and $q = (q_k^{k-1})_{k=2}^n$. Since $q_k^{k-1} = W_{k-1}'/
W_{k-1}$, we have $\frac{Dq_k^{k-1}}{D\phi} =
\partial\left(\frac1{W_{k-1}} \frac{DW_{k-1}}{D\phi}\right)$. Thus, the
equality of the evolutions \eq{maine1} and \eq{maine100} will be proved once we
show that
$$
\Omega^{-1}S\left(\frac1W\frac{DW}{D\phi}\right)\,\Phi\,(\Id+A)\, \I =
-\frac1n J \left(\frac{Du}{Dv}\right)^{\ast} \widehat{\frac{\delta \H}{\delta
u}},
\Eq{maine2} 
$$
where by $\frac1W\frac{D W}{D \phi}$ we mean the matrix
whose $(i, j)$ entry is given by $\frac{1}{W_i} \frac{D W_i}{D
\phi_j}$. Straightforward multiplication of matrices shows that \eq{maine2}
becomes 
$$
\frac1W\, \frac{D W}{D \phi}\, \Phi\,(\Id + A)\, \I = R 
\left(\frac{Du}{Dy}\right)^{\ast} \widehat{\frac{\delta \H}{\delta u}},
\Eq{maine3}
$$
where 
$$
R = -\frac1n\,S^{-1}\,\Omega\,J\,\Omega^t=
\pmatrix{{-1}&1&0&0&\dots &0\cr
{-2}&1&1&0&\dots &0\cr
-3&1&1&1&\dots &0\cr
\vdots &\vdots &\ddots &\ddots &\ddots &\vdots \cr
{-(n-2)}&1&\dots &1&1&1\cr
{-(n-1)}&1&\dots &1&1&1\cr
}.
$$
We conjecture \eq{maine3} to be true whenever 
$$
\frac{\delta \H}{\delta u} =
T \M\, \I,
$$
where
$$
T = 
\pmatrix{0&\dots &0&0&0&0&1\cr
0&\dots &0&0&0&1&\partial \cr
0&\dots &0&0&1&\bin21\partial &\partial ^2\cr
0&\dots &0&1&\bin31\partial &\bin32\partial ^2&\partial ^3\cr
\vdots &\adots &\adots &\adots &\adots &\adots &\vdots \cr
0&1&\bin{n-3}1\partial &\bin{n-3}2\partial ^2&\bin{n-3}3\partial ^3&\dots
&\partial ^{n-3}\cr 
1&\bin{n-2}1\partial &\bin{n-2}2\partial ^2&\bin{n-2}3\partial^3&\dots
&\bin{n-2}{n-3}\partial ^{n-3}&\partial ^{n-2}\cr }
$$
and $\M$ is a certain upper triangular matrix of the form
$$
\M = \pmatrix{ 1&0&m_1^1&m_1^2&\dots&m_1^{n-3}\cr 
0&1&0&m_2^2&\dots&m_2^{n-3}\cr 
\vdots&\ddots&\ddots&\ddots&\ddots&\vdots\cr
0&\dots&0&1&0&m_{n-3}^{n-3}\cr 
0&\dots&0&0&1&0\cr 
0&\dots&0&0&0&1},
$$
whose matrix elements $m_i^j$ are all functions of the coefficients $u_i$
and their derivatives. On the other hand, if $H_u =
\sum_{k=1}^n h_k \partial^{-k}$, the vector $(h_1,\dots,h_{n-1})$ is
easily seen to be related to the gradient of $\H$ through the matrix $T$,
excatly the same way $\M\,\I$ is. (The coefficient $h_n$ of $H_u$ is determined
by the other coefficients, from the condition that the associated Hamiltonian
vector field
$V_{H_u}$ be tangent to $\An$.) That is, \eq{maine3} {\it will hold
provided a certain linear combination of $\I$ with differential
invariant coefficients coincides with the coefficients $(h_1,\dots,h_{n-1})$ of
the pseudo-differential operator
$H_u$ defining the evolution of $u$}.

One can see this relation between $\I$ and $H_u$ from a different point of
view. Any relative invariant is the product of the particular solution $\mu$
of \eq{sym2} given by
\eq{mat2}, times an invertible matrix of
differential invariants, such as $\M$. That is, we conjecture that one can find
a relative invariant of the form $\tilde\mu =
\Phi\,(\Id+A)\,\M^{-1}$ such that  the evolution $\phi_t = \tilde\mu\,\I$ is
equivalent to the Adler--Gel'fand--Dikii evolution whenever $\I$ equals the
coefficients of $H_u$. This gives a {\it Hamiltonian interpretation
of $\sL n$ differential invariants}.

Finally, \eq{maine3} becomes the following equality of matrices:
$$
\frac1W\, \frac{D W}{D \phi}\, \Phi\,(\Id + A)  = R 
\left(\frac{Du}{Dy}\right)^{\ast} T \M.   \Eq{maine}
$$
Let us analyze this equation. The matrix $\frac1W\,\frac{D W}{D \phi}\,\Phi$
is easily calculated to have as $(i,j)$ entry the expression
$\sum_{r=1}^i\sum_{s=0}^r
\bin rs q_{1\, \dots\, r+j-s\,\dots\, i}\, \partial^s$, where
$r+j-s$ is in the
$r^{\rm th}$ place. Thus, the left-hand side of \eq{maine} does not represent
a major problem. With respect to the right-hand side, we can write this
expression in terms of $q$'s. There are old formulas, \rf{W}, relating $u$'s
to $q$'s which, in our notation, become
$$
u_m = \sum_{i=0}^{n-m} (-1)^{n-m-i} \bin{m+i}{i}
\Lambda_i\,q_n^{m+i},
\qquad 0\le m\le n-2,
\Eq{Wil1}
$$
where $q_n^0=0$ by definition, $\Lambda_0 = 1$, $\Lambda_1 = \frac1n
q_n^{n-1}$, and $\Lambda_i$ is given by the following recurrent formula
$$
\Lambda_i = \sum_{k=0}^{i-1} \bin{i-1}k \Lambda_{k}\,
(\Lambda_1)^{(i-1-k)}.
\Eq{Wil2}
$$
In particular, observe that the $\Lambda_i$'s are all functions of $q_n^{n-1}$
and its derivatives. Using formulas similar to these and
Lemma \th{form} skillfully enough, one should expect to be able (although this
is by no means trivial) to simplify that part of the equation. The main
trouble lies with the expression of
$\M$. For
$n=2$ and $3$
$\M$ is the identity. For $n=4$
$$
\M=\pmatrix{1&0&-\frac12u_2\cr 0&1&0\cr 0&0&1},
$$
and for $n=5$
$$
\M=\pmatrix{1&0&-\frac7{10}u_3&\frac35u_2-u_3'\cr 0&1&0&-\frac25u_3\cr
0&0&1&0\cr 0&0&0&1}.
$$
But for higher dimensions $\M$ involves more complicated expressions of the
coefficients $u$ in a fashion we were not able to decipher. Hence the
difficulty of proving \eq{maine} in the general case. We were also unable to
show the existence of $\M$ in the general case. On the other hand, one can use
all the hints given here to attack a fixed dimension, and we did so up to
$n=6$, finding the value of $\M$ straight from the equation itself. The main
problems is the choice of variables; in fact, the goal would be to find
a different set of variables making the equivalence between the
Adler--Ge'lfand--Dikii and the $\SLn$ invariant evolutions totally
transparent.

\vfill\break
\parskip=6pt plus 1pt \parindent=0pt \frenchspacing
{\Bf References}
\medskip
\[AH] Ackerman, M., and Hermann, R., {\it Sophus Lie's 1884 Differential
Invariant Paper,} Math Sci Press, Brookline, Mass.~(1975).

\[A] Adler, M., {\it On a Trace Functional for Formal Pseudo-differential
Operators and the Symplectic Structure of the KdV,} Inventiones Math.~{\bf
50}, 219--48~(1979).
 
\[DS] Drinfel'd,  V.G. and Sokolov, V.V., {\it Lie Algebras and Equations of
KdV Type,}  J. of Sov. Math.~{\bf 30}, 1975--2036~(1985).

\[GD] Gel'fand, I.M. and Dikii, L.A.  {\it A family of Hamiltonian structures
connected with integrable nonlinear differential equations,} in I.M.
Gel'fand, {\sl Collected papers,} Vol.~1, Springer--Verlag~(1987).

\[H] Halphen, G.-H., {\it Sur les invariants diff\'erentiels,}, in {\sl
Oeuvres,} Vol.~2, Gauthier--Villars, Paris~(1913).

\[KW] Kupershmidt, B.A. and Wilson, G., {\it Modifying Lax Equations
and the Second Hamiltonian Structure,} Inventiones Math.~{\bf 62},
403--36~(1981). 

\[L] Lie, S., {\it \"Uber Differentialinvarianten,} in {\sl Gesammelte
Abhandlungen,} Vol.~6, B.G. Teubner, Leipzig~(1927); see \rf{AH} for an
English translation.

\[OST] Olver, P.J., Sapiro, G., and Tannenbaum, A., {\it Differential invariant
signatures and flows in computer vision: a symmetry group
approach,} in {\sl Geometry-Driven Diffusion in Computer Vision,} B.M. ter
Haar Romeny, ed., Kluwer Acad. Publ., Dordrecht, The Netherlands~(1994).

\[O1] Olver, P.J., {\it Applications of Lie Groups to Differential
Equations,} Second Edition, Springer--Verlag, New York~(1993).

\[O2] Olver, P.J., {\it Equivalence, Invariance and Symmetry,}
Cambridge University Press, Cambridge, UK~(1995).

\[OK] Ovsienko, V.Yu. and Khesin, B.A., {\it Symplectic leaves of the
Gel'fand--Dikii brackets and homotopy classes of non-degenerate curves,} Funct.
Anal. and Appl.~{\bf 24} n.~1, 33--40~(1990).

\[T] Tresse, A., {\it Sur les invariants diff\'erentiels des groupes continus
de transformations,} Acta Math.~{\bf 18}, 1--88~(1894).

\[W] Wilczynski, E.J., {\it Projective differential geometry of curves and
ruled surfaces,} B.G. Teubner, Leipzig~(1906).

\[Wi] Wilson, G., {\it On the antiplectic pair connected with the
Adler--Gel'fand--Dikii bracket,} Nonlinearity~{\bf 5}, 109--31~(1992).

\bye